\documentclass[%
 reprint,
 amsmath,amssymb,
 aps,
]{revtex4-2}

\usepackage{graphicx}
\usepackage{dcolumn}
\usepackage{bm}

\begin{document}

\preprint{APS/123-QED}

\title{Graph Variational Autoencoder for Detector Reconstruction and Fast Simulation in High-Energy Physics}

\author{Ali Hariri}
 \email{aah71@mail.aub.edu}
\affiliation{%
 American University of Beirut
}%
\author{Darya Dyachkova}
\affiliation{Minerva Schools at KGI}%

\author{Sergei Gleyzer}
\affiliation{%
Department of Physics and Astronomy, University of Alabama
}%

\date{\today}

\begin{abstract}
  Accurate and fast simulation of particle physics processes is crucial for the high-energy physics community. Simulating particle interactions in the detector is both time consuming and computationally expensive. With a proton-proton collision energy of 13 TeV, the Large Hadron Collider is uniquely positioned to detect and measure the rare phenomena that can shape our knowledge of new interactions. The High-Luminosity Large Hadron Collider (HL-LHC) upgrade will put a significant strain on the computing infrastructure and budget due to increased event rate and levels of pile-up. Simulation of high-energy physics collisions needs to be significantly faster without sacrificing the physics accuracy. Machine learning approaches can offer faster solutions, while maintaining a high level of fidelity. We introduce a novel graph generative model that provides effective reconstruction of LHC events on the level of calorimeter deposits and tracks, paving the way for full detector level fast simulation.
\end{abstract}

\maketitle


\section{Introduction}
 Accurate simulations of high-energy physics (HEP) experiments play an important role in various tasks, including optimization of detector geometry, optimization of physics analyses, and searches for new phenomena beyond the Standard Model (SM) \cite{morais2020phenomena, kuno2001muon, beacham2019physics}. Typically, billions of events are simulated by HEP experiments across the many SM processes and sought-after hypothetical signals to achieve the desired performance required by the Large Hadron Collider (LHC) experiments to test the SM predictions. The uncertainty in these various analyses is directly related to number of events in the simulated samples, which leads to considerable amount of computational budget spent on simulations, frequently on a scale comparable to data processing. 
 
 The next generation of HEP experiments, such as the High-Luminosity Large Hadron Collider (HL-LHC) \cite{Apollinari:2015bam} is projected to reach a peak instantaneous luminosity of up to $7.5\times10^{34}$~cm$^{-2}$s$^{-1}$, a factor of four increase beyond LHC Run 2. The HL-LHC goal is to collect an integrated luminosity of at least $3000$~fb$^{-1}$ in ten years of operations. At HL-LHC, the number of independent proton-proton collisions per bunch crossing (`pileup') is expected to reach a mean value of 200, a very large increase above the Run 2 mean of $\sim40$. 
 
 Advanced computational paradigms, including advanced machine learning techniques and dedicated hardware that accelerates their application for detector simulation and event reconstruction, will be necessary to attain the main physics goals of the HL-LHC due to significant challenges posed by increased levels of pile-up and the rarity of sought-after signals \cite{CWP}. 
 
 A typical approach to fast detector simulation involves either a simplified geometry or a simplified model of interactions of particle with the detector material \cite{fastsimatlas}. This is adequate for basic simulation tasks that do not require high levels of fidelity. However, for LHC experiments that rely on detailed descriptions of modern calorimeters and tracker systems, the trade-off between speed and accuracy becomes too large for practical applications. 
 
 Frequently, fast simulation approaches parametrize the detector response \cite{Yamanaka_2011, delphes}. The idea behind such group of methods is to define a phase space based on the characteristics of the physical systems and then approximate the underlying probability by repeatedly sampling from that space. The simulation is constrained by the ground truth, while the quality of the samples depends on the definition of the phase space \cite{deeplearn4lhc}. A technique of "frozen showers" is sometimes used in calorimetry simulations, where the detailed simulated calorimeter responses are stored in an offline library for subsequent use \cite{fastsimatlas, Yamanaka_2011}. Parametrized models, such as Delphes \cite{delphes}, are faster than full simulation, but again due to simplifying assumptions about the detectors suffer from a loss in fidelity compared to full simulation, and are not meant to be used for advanced detector studies and analyses. 
 
 Probabilistically, the reconstruction of particle collision events is modeled using the equation: 
 \small
\begin{align}
\begin{split}
   p(\textrm{r-particles}|\theta)&=\int
   R(\textrm{r-particles}|\textrm{particles})H(\textrm{particles}|\textrm{partons}) \\
&\times P(\textrm{partons}|\theta)\textrm{ } d \textrm{particles}\textrm{ }d\textrm{partons}
\end{split}
\end{align} 
\normalsize
where P is the probability density of observing a set of reconstructed objects given a point in the parameter space, H refers to the hadronization process where mapping from the parton to the particle level occurs and R(particles) is the detector response \cite{leshouches}. The latter is estimated using Monte Carlo-based full simulation or parametric methods. 


\section{Related Work}
The current forward simulation models, such as Geant \cite{Geant} are accurate but difficult to parallelize and do not scale well with the rising computational complexity. Recent developments and successes of machine learning in high-energy physics (HEP) \cite{CWP, dl4hep, Cogan_2015, cnn4hep, kasieczka2019machine, denby1988neural, du2021deep} have opened new possibilities for fast simulation. An active area of current research is fast detector simulation based on generative models \cite{butter2020generative}, in particular Generative Adversarial Networks (GANs) \cite{goodfellow2014generative} and Variational Autoencoders (VAEs) \cite{vae}. For example, the use of GANs for fast simulation was first proposed in \cite{de_Oliveira_2017} and further developed for many other applications \cite{Di_Sipio_2019, Paganini_2018, Paganini_acce, Ghosh:2680531}.
Similarly, VAEs have been proposed as another alternative to for fast simulation \cite{Ghosh:2680531}. In \cite{gettinghigh}, a combination of VAEs and GANs was used to simulate electromagnetic showers. These deep learning-based approaches allow a faster inference as only a forward pass through a neural network is required, while the fidelity levels have improved over time \cite{gettinghigh}.

Graph neural networks have been recently applied to a number of applications in high-energy physics \cite{GNNHEP}. In contrast to other machine learning methods, graph representation learning is able to handle non-Euclidean geometry and irregular grids \cite{bruna2013spectral},  introduce relational inductive bias into data-driven learning systems \cite{battaglia2018relational} and encode physics knowledge in graph construction. Particle collision events can be described by point clouds mapped into graphs. In \cite{particlecloud},  a deep learning architecture with similar properties as DGCNN in \cite{wang2019dynamic}, ParticleNet, is applied to the challenges of top jet tagging \cite{kasieczka_gregor_2019_2603256} and quark/gluon tagging \cite{energyflownetwork}. Other applications of graph neural networks in high-energy physics include pileup mitigation \cite{martinez2019pileup}, particle reconstruction \cite{ju2020graph, gravnet} and particle tracking {\cite{ju2020graph}}. 
The Gated Graph Neural Network (GGNN) architecture for pile-up mitigation \cite{martinez2019pileup} has achieved good results compared to baselines, such as PUPPI \cite{Bertolini_2014}, GRU \cite{chung2014empirical} and the SoftKiller algorithm \cite{Cacciari_2015}. Interaction Networks from \cite{battaglia2016interaction} have also been applied to particle track reconstruction \cite{dezoort2021charged}. The data from the barrel region of the detector is pre-processed by assigning cylindrical coordinates $(r,\phi,z)$ as node features and the differences between them $(\Delta\eta, \Delta\phi)$ as edge features. The model resulted in an overall relative efficiency of $95\%$ for track finding. 


In this study, we propose a novel Graph Variational Autoencoder model that combines elements of VAEs and geometric deep learning \cite{gnn} in order to accurately learn a compressed representation of the data for reconstruction of high-energy physics events. More specifically, we develop a novel generative model to learn the representations of events in high dimensional space to embed such generative models into a full simulator. We further develop spatial graph convolutional layers \cite{spatialgcn} to learn the properties of the graph-like structures and spectral clustering layers to compress these graphs into smaller, more representative nodes.
\section{Geometric Deep Learning}

\subsection{Definition}

We briefly introduce the basics of geometric deep learning relevant to this study. A graph is denoted by $G= (V,E,A)$ where V is the set of vertices composing the graph, E the set of edges connecting these vertices, and A being an N x N adjacency matrix, where N is the total number of nodes in the graph. Associated with each vertex V is a set of features describing it. These are given in a feature vector $X \in R^{N \times D}$ with D being the number of features per node.Let $v_i \in V$ denote a vertex and $e_{ij}=(v_i,v_j) \in E$ an edge connecting this vertex to a neighbor $v_j$. $A_{ij}$ then denotes the value $w_{ij}$ at the $j^{th}$ column of the $i^{th}$ row of the adjacency matrix. In the case of a weighted adjacency matrix, $w_{ij}$ represents the weight of an edge connecting two vertices, indicating the effect presented by $v_j$ on $v_i$’s features as compared to other neighbors. An un-weighted adjacency matrix is characterized by a  $w_{ij}$ value equal to 1 if a connection exists between two vertices, and a value of 0 otherwise. The total sum of neighbors for a node $v_i$ represents the degree of this node and is given by $\sum_{j=0}^{n}w_{ij}$ where n is the number of neighbors. At this stage, we distinguish two types of graphs: A directed graph is one where the edges unidirectionally point from one vertex to the neighboring one. As a result, it is likely that $w_{ij} \neq w_{ji}$ in such a case. In contrast, connectivity and edge features go both ways in an un-directed graph hence $w_{ij}= w_{ji}$ [21].

\subsection{Graph Networks}
Early work on graph analytics used Fourier transforms and other mathematical operations to transfer to graph-based signals as shown in \cite{Chen_2015}. For instance, convolution operations that have shown great success with CNNs have been combined with spectral graph theory as a first attempt to convolve non-structured data \cite{bruna2014spectral}.
In \cite{kipf2017semisupervised}, spectral graph convolution, a graph-based convolution operation with origins in graph signal processing and graph theory is introduced. An $N \times N$ adjacency matrix A defines the topology of the graph, i.e. the edge connections between the nodes. A feature vector X of shape $N \times F$ represents the node features of dimension F. Spectral graph convolution is performed by first obtaining the normalized graph Laplacian whose eigenvectors form the basis of the orthonormal space to which the graph Fourier-transformed input signals are projected. Next, the graph convolution operation is defined as a dot product between the transformed graph signals and a spectral kernel with trainable parameters \cite{Wu_2021}. The latter can be expressed as a truncated expansion of Chebyshev polynomials as done in \cite{hammond2009wavelets}.



\begin{equation}
    g_{\theta}(\Lambda)=\sum_{k=0}^{K-1}\theta_k T_k (\Tilde{\Lambda})
\end{equation}
where $\Tilde{\Lambda}= \frac{2}{\lambda_{max}}\Lambda - I_n$, $\lambda_{max}$ being the largest eigenvalue in $\Lambda$.\\
\\
The kernel $g_\theta$ is applied to the diagonal matrix of eigenvalues. $\Theta$ is a vector containing Chebyshev coefficients to reduce the number of parameters, T contains Chebyshev polynomials and K is the order of neighborhood away from a given node to be covered. 

\subsection{Spatial graph convolution}
Spectral-based graph convolution architectures such as GCN \cite{kipf2017semisupervised}  and CayleyNets \cite{levie2018cayleynets} have shown notable success in several tasks such as image classification and citation networks . Nevertheless, these models are associated with high computational cost with relatively high time and memory complexity due to operations such as the computation of the diagonal degree matrix. More recent approaches consider spatial-based convolution operations whereby the node features are updated with message passing between the neighbouring nodes over k neighborhoods. In \cite{gilmer2017neural}, Message Passing Neural Networks (MPNNs) operate on the QM9 dataset for chemical properties prediction, achieving low mean absolute error. Interaction Networks were also used to study more dynamic phenomena in simulation scenarios \cite{battaglia2016interaction}. Such methods require message passing between the nodes over K iterations in order to map the initial node features $h_0$ at timestep 0 to $h_t$ at timestep t. A general formulation of a message passing process is given by: 
\begin{equation}
    h_v^k=U_k(h_v^{(k-1)} , \sum_{u \in N(v)} M_k(h_v^{(k-1)},h_u^{(k-1)},x_{vu}^e))
\end{equation}

where $U_k(.)$ and $M_k(.)$ are functions with trainable parameters \cite{Wu_2021}. By comparison, spatial convolution architectures are more efficient in time and memory complexity than spectral-based convolution as shown in \cite{Wu_2021}. A wide variety of message passing GNNs exist, some of which perform random neighbourhood sampling for faster operations on smaller neighbor samples such as GraphSAGE \cite{sagegraphs} while others add attention mechanisms to the architecture resulting in better accuracy \cite{veli2018graph}.

\section{Data and Pre-Processing}
\begin{figure}
    \centering
        \includegraphics[scale=0.28]{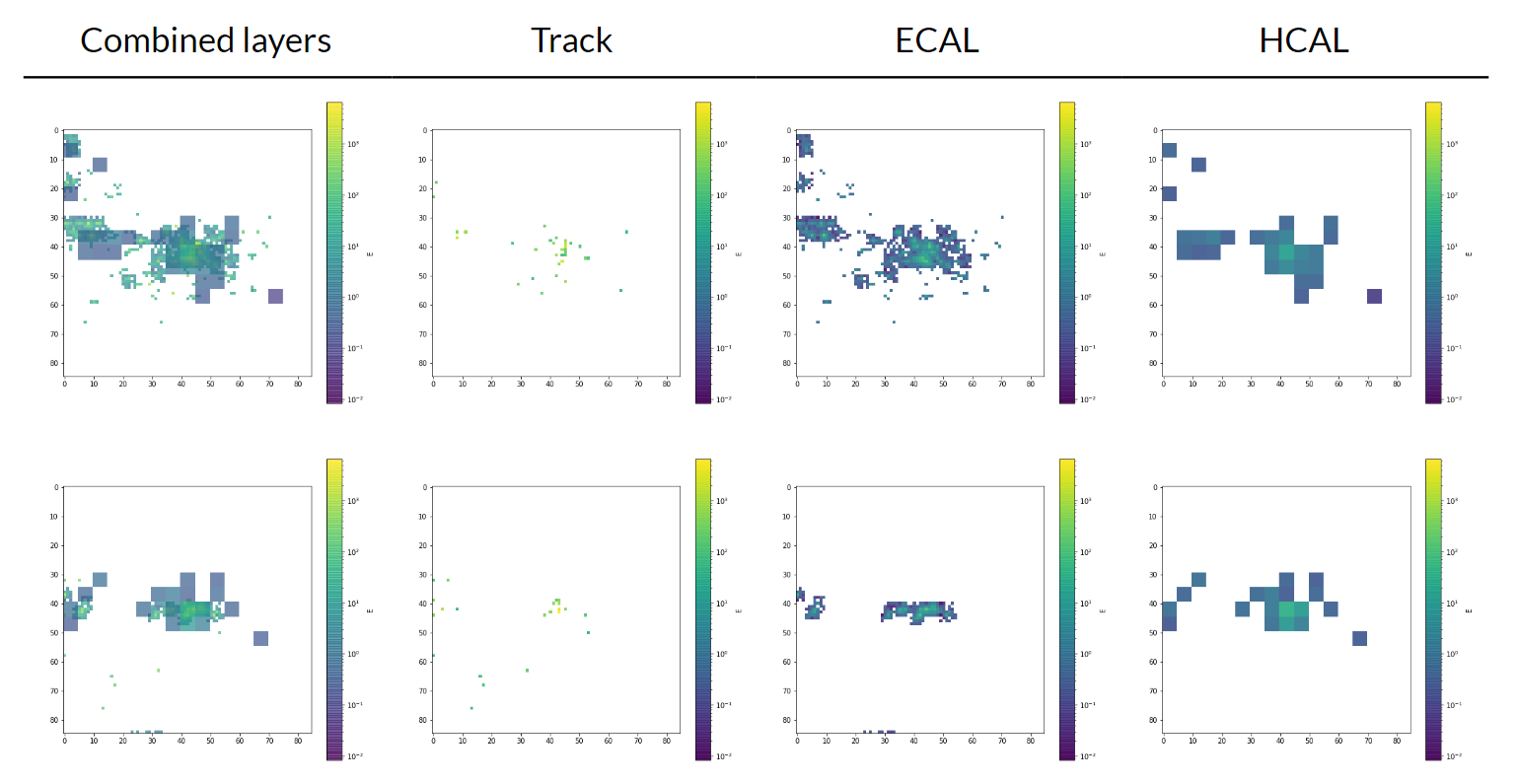}\\
    \caption{The visualization of the energy deposits in ECAL and HCAL calorimeter layers for two distinct $ttbar$ events (top and bottom). The tracks layer shows the tracks projected to the ECAL surface. The leftmost image is the combined one from all the three layers. } 
    \label{fig:figs}
    \vspace{-0.5cm}
\end{figure}

In this work, we focus on top quark pair events available on the CERN Open Data Portal \cite{cmsopendata} that are originally simulated using the Pythia 6 generator \cite{sjostrand2006pythia}. The input data consists of around 30000 samples, each of which is characterized by three sub-detector channels: Tracks (TK), Electromagnetic Calorimeter (ECAL) and Hadronic Calorimeter (HCAL). For HCAL and ECAL subdetectors, the inputs are reconstructed calorimeter hits. The Track inputs consist of projected tracks to the ECAL surface. To visualize the events in different sub-detectors, we display them on a mesh with 85 x 85 segmentation (Fig. \ref{fig:figs}), shown at ECAL resolution. A full description of the pre-processing steps used in this work can be found in \cite{e2eclassification, Andrews_QGC}. The features of interest are $i\eta$ and $i\phi$ locations and reconstructed hit energies. 


\section{Graph Representation of Events}

We use the k-nearest neighbour algorithm to connect each node representing a reconstructed hit in a detector cell to k neighbouring cells closest in Euclidean distance given by $\sqrt{(x-x_i)^2+(y-y_i)^2}$. Therefore, the reconstructed detector hits are mapped into nodes that contain 3 features: x and y locations and the reconstructed energy of the hit. We next learn the high dimensional representation of the nodes using a Graph Variational Autoencoder (GVAE), a geometric deep learning architecture that learns the graph embeddings of the non-Euclidean data in a latent space \cite{gvae}.



\begin{figure}
    \centering
        \includegraphics[scale=0.3]{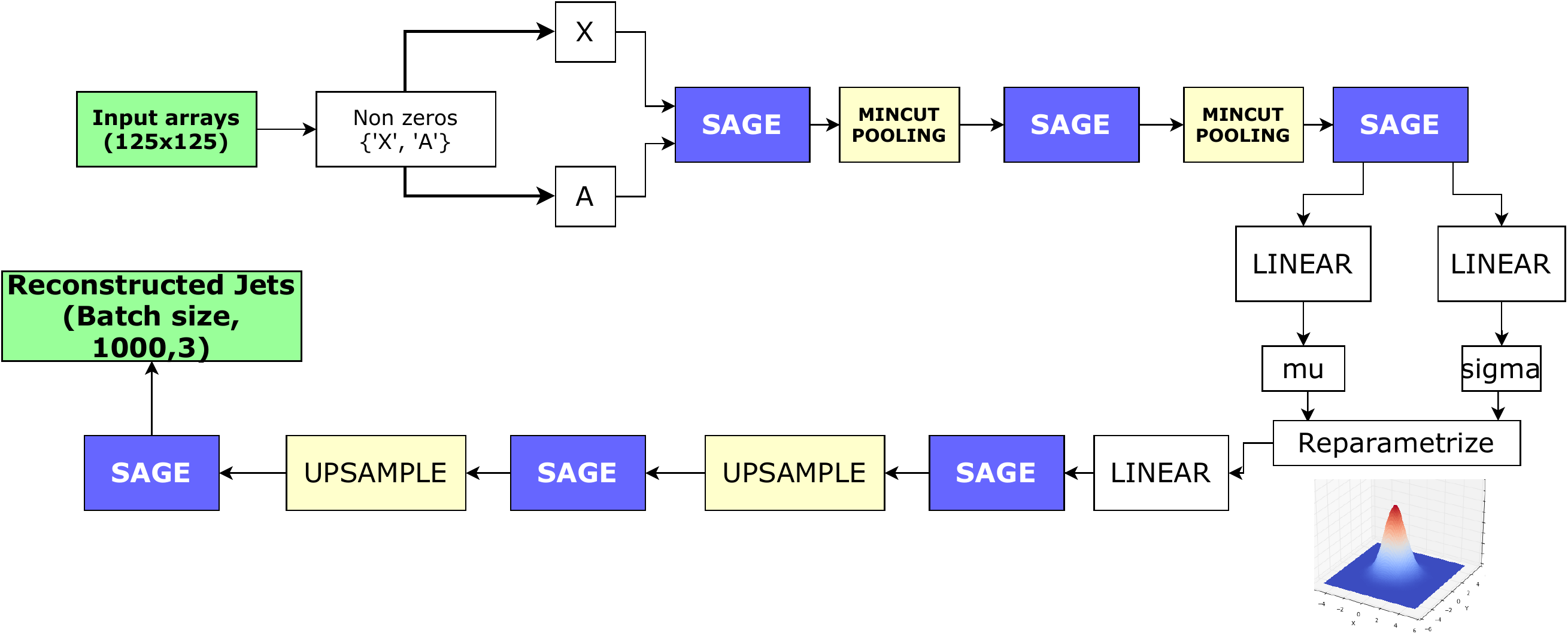}\\
    \caption{Model architecture of the Graph Variational Autoencoder showing GraphSAGE layers and pooling blocks}
    \label{fig:arch}
\end{figure}

GraphSAGE \cite{sagegraphs} is an effective spatial graph convolution architecture and we use it to learn the node properties. GraphSAGE is an inductive learning framework that considers nodes of similar semantics or features to have nearby positions in latent space. Feature information is aggregated from k higher-order neighbourhoods towards each central node, and we refer to k as the depth of the message passing layer. This technique is at the core of the encoder-decoder layers to learn the representations of particle features into a compressed high dimensional space. The compression and up-sizing of the graphs in latent space are accomplished with the MinCut spectral pooling approach defined in \cite{graphpooling}. 


\begin{equation}
X^{rec}=SX^{Pooled}; A^{rec}=SA^{Pooled}S^T
\end{equation}
where S is a learned cluster assignment matrix similar to the one defined in \cite{graphpooling}. A pictorial representation of the GVAE model is given in Figure \ref{fig:arch}. 


\section{Results}
 We trained the Graph Variational Autoencoder model on a Tesla T4 GPU on Google Colaboratory \cite{inbook} to learn the representations of top quark-initiated jets. Figure 3 provides a visual comparison between the original fully-simulated events in the top row and the GVAE reconstructed events in the bottom row. A visual assessment shows a high degree of similarity between the GVAE reconstruction and the original simulation. Using the Earth Mover Distance metric \cite{EMD} to quantify the cost of displacing the point clouds of the original events to the reconstructed events (Figure \ref{fig:emd}) we visualize the performance of the sample. The EMD metric refers to the "work" required to convert one event $\varepsilon$ into another $\varepsilon'$ by moving energy between particles $i$ and $j$ in the two events. It is described by the following equation: 

\begin{align}
    EMD(\varepsilon, \varepsilon') &= \min_{\big\{f_{i,j}\big\}} \sum_{ij}f_{ij} \frac{\theta_{ij}}{R} + \Big|\sum_i E_i - \sum_j E_{j}'\Big|\\
    f_{ij} \geq 0, \hspace{0.5cm} \sum_j f_{ij} &\leq E_i, \hspace{0.5cm} \sum_i f_{ij} \leq E_j', \hspace{0.5cm} 
    \sum_{ij} f_{ij} = E_{min}
    \end{align}

The EMD distribution indicates good agreement across the full sample of events. The model achieves an average inference time of 0.13088 seconds over several runs on a batch size of 64, which is an order of magnitude below full reconstruction times for simulation of top quark pair events, typically of the order of tens of seconds, while maintaining a very high level of fidelity. 

\begin{figure}
\begin{center}
    \includegraphics[scale=0.40]{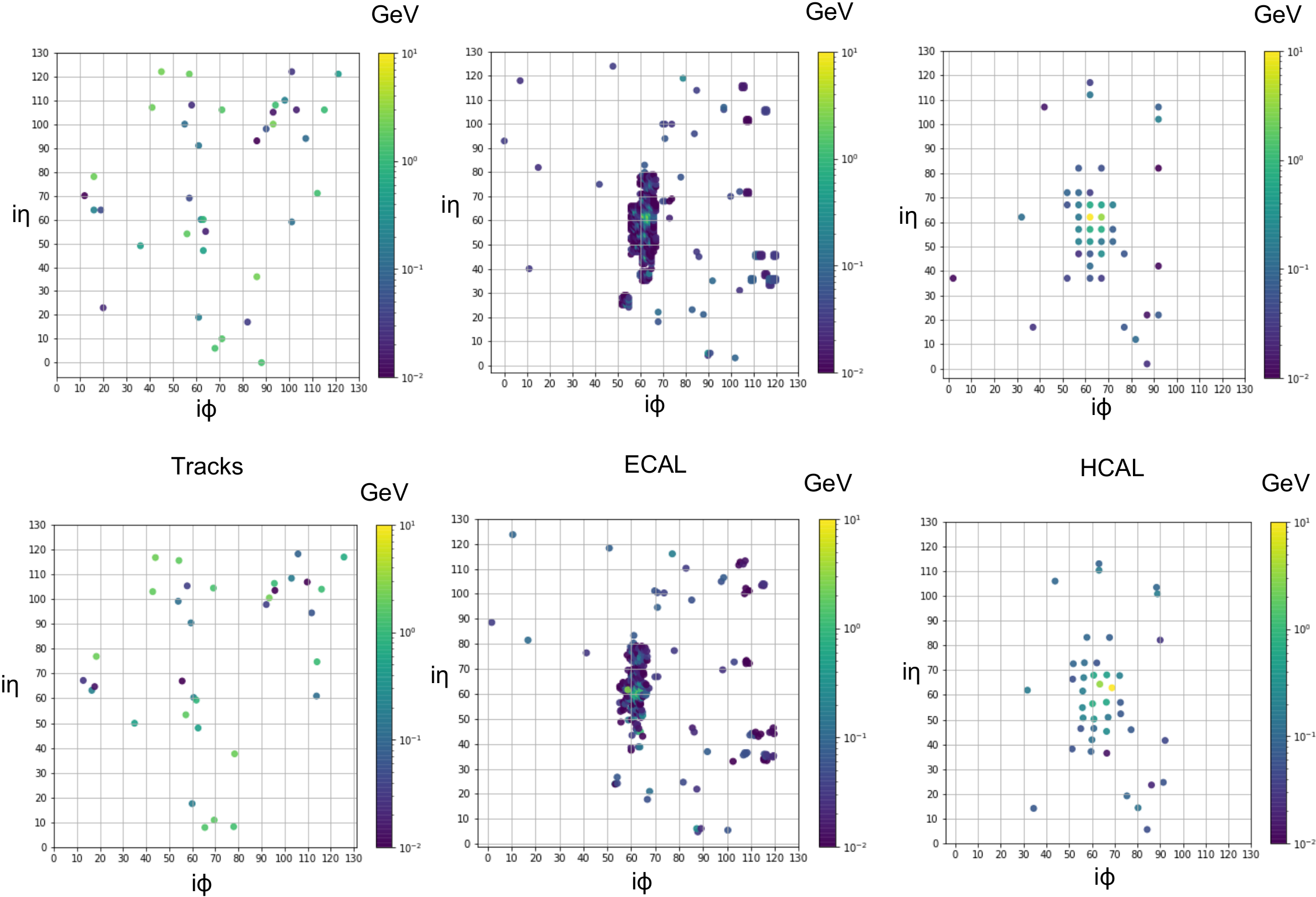}
    \caption{Original simulated top quark initiated jet (top) compared to the GVAE-reconstructed jet (bottom) in each of the three channels. The energy range is log-scaled for better visualization.} 
\end{center}
\label{fig:compare}
\end{figure}

\begin{figure}
\begin{center}
        \includegraphics[scale=0.5]{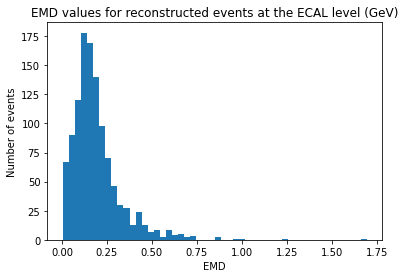}
    \caption{Earth Mover Distance distribution as described in Ref. \cite{EMD}. Low values of EMD indicate good agreement.}
        \label{fig:emd}
\end{center}
\vspace{-1cm}
\end{figure}

\subsection{Multi-GPU Scaling}

We also investigate how the GVAE model scales on multiple GPU devices. We first profile the code to monitor the process load taking place in CPUs and GPUs. To avoid a potential data-loading bottleneck, we generate batches of graphs on the spot in the CPU prior to sending them to the GPU. This change has reduced the training time by $50\%$. After performing baseline training on a single Tesla V100 GPU, we scaled the model with Horovod \cite{sergeev2018horovod}, a deep learning library for distributed training of deep learning models that supports PyTorch \cite{paszke2019pytorch}, TensorFlow \cite{abadi2016tensorflow} and MXNet \cite{chen2015mxnet}. This framework facilitates distributed training across multiple GPUs using the Message Passing Interface (MPI) and takes as input the $size$ parameter referring to the number of processes P in training, followed by the $Rank$ parameter unique to process and numerated from 0 to P-1. Finally, $Local Rank$ indicates the unique process ID within each device (0 to $N_{GPUS}$).

We used NVIDIA's Automatic Mixed Precision (AMP) library \cite{ampnvidia} that enables mixed precision training of deep learning models with support to distributed parallel training. This approach has the potential to boost the speedup of model training by enabling the support of FP16 operations throughout the model's layers. This step offers several benefits such as the reduction in required memory for FP16 as compared to 32-bit floating points, faster data transfer and linear algebra operations with lower precision formats. Using multi-GPU scaling methods allows us to additionally speed-up the performance during training and inference, something that is very challenging to accomplish with present simulation methods.

\begin{figure}[h]
     \centering
    \includegraphics[scale=0.43]{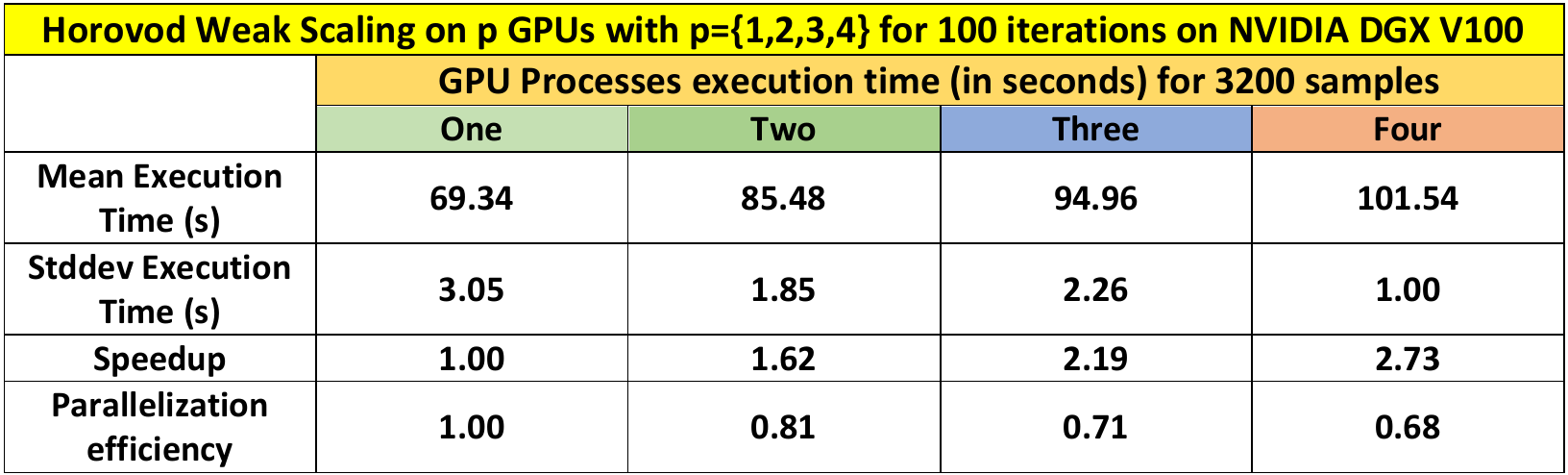}
       \caption{Table showing scaling efficiency }
    \includegraphics[scale=0.22]{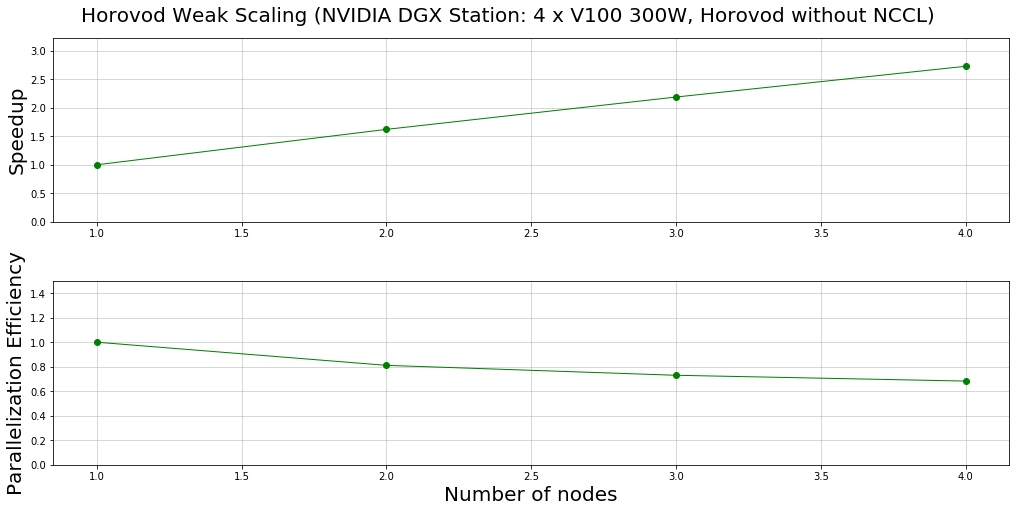}
    \caption{Comparison plot of how our model scales to multiple GPUs}
    \label{fig:gpuscale}
    \end{figure}

We train the GVAE model on a cluster using Volta V100 GPUs with 16 GB of RAM. Following profiling and code optimization for enhanced CPU performance, the training is scaled on multiple GPUs using the Horovod library. We compare the results to the training on a single GPU with a batch size of 32 for 100 iterations, i.e a total of 3200 graph samples (Figure \ref{fig:gpuscale}). 

We notice an increase in the performance with an increase in the GPU devices used. To calculate the resulting speedup, we take as reference the Mean Execution Time (MET) resulting from training on one GPU. For $N_{GPUs}=2$, MET is 85.48 seconds. Bearing in mind that we are training twice the amount of data overall, we get MET per GPU = 42.74 seconds. The resulting speedup for $N_{GPUs}=2$ is given by 
\begin{equation}
    \frac{MET_{single GPU}}{MET per GPU using {N_{GPUs}}}=\frac{69.34}{42.74}=1.62
\end{equation}
Applying the same speedup calculation, going from 2 to 4 GPUs, we obtain a speedup factor of 1.62, 2.19 and 2.73, respectively. We conclude that the GVAE model scales well on multiple GPUs using Horovod, a useful result for future work and similar graph generative models. However, there is still some loss in scaling performance due to parallelization efficiency that can be explained by an inefficient communication between the GPUs once parallel batch training has been done. This can be overcome by programming the kernel directly. Further optimization is deferred to future work.


\section{Conclusion}
In this work we explore a novel graph-based generative architecture for learning the representation of high-energy collision events. The use of graph representation aligns well with the sparsity of the particle detectors hits. Through spatial convolution the GVAE model was able to learn the interactions between hits in the detector, while mincut pooling sequentially compressed it, preserving the most representative nodes in latent space. Finally, a trained decoder can be used to upsample the compressed vectors to generate new events for fast detector simulation. We additionally benchmarked the GVAE model on single and multiple GPUs, obtaining low latency and efficient multi-GPU scaling performance.

\section{Acknowledgments}
We would like to thank Michael Andrews, Bjorn Burkle and Daria Morozova for useful discussions. We additionally thank Mozhgan Kabiri Chimeh from NVIDIA in addition to Giuseppe Fiameni and Christian Hundt from the NVIDIA AI Technology Center (NVAITC) for support during a GPU hackathon that enabled us to study multi-GPU scaling. S. G. has conceived the overall idea, contributed to algorithmic design and co-wrote the manuscript. A.H developed the graph autoencoder model presented in this work for jet reconstruction. D. D. has developed a non-graph autoencoder that has informed the group regarding the specificity of the latent space of the data features, as well as wrote the dataloading, partial preprocessing, and Earth Mover Distance evaluation scripts. A. H. was partially
supported by an IRIS-HEP fellowship through the U.S. National Science Foundation (NSF) under Cooperative Agreement OAC-1836650. A. H. is also a participant in the Google Summer of Code 2021 Program. S. G. is supported by U.S. Department of Energy (DOE) under Award No. DE-SC0012447. 

\newpage
\bibliographystyle{unsrt}
\bibliography{references.bib}
\end{document}